\documentclass{aa}  
\usepackage{graphicx}
\usepackage{txfonts}
\begin{document}
   \title{Radio continuum imaging of the R CrA star-forming region  
with the ATCA} 
 
   \author{O. Miettinen\inst{1}, S. Kontinen\inst{1}, J. Harju\inst{1}, 
           \and J. L. Higdon\inst{2}} 
 
   \offprints{O. Miettinen} 
 
   \institute{Observatory, P.O. Box 14, 00014 University of Helsinki, Finland\\ 
              \email{oskari.miettinen@helsinki.fi} 
         \and  
              Dept. of Physics, Georgia Southern University, Statesboro, 
              GA 30460-8031} 
 
   \date{Received 3 January 2008 / Accepted 28 April 2008} 
 
% \abstract{}{}{}{}{}  
% 5 {} token are mandatory 
\authorrunning{Miettinen et al.} 
\titlerunning{Imaging of R CrA star-forming region}

  \abstract
   {}
   {The aim of this study is to investigate the nature of radio 
sources associated with young stellar objects (YSOs) belonging to the 
R CrA cluster. By combining the centimetre radio data with the wealth 
of shorter wavelength data accumulated recently we wish to refine
estimates of the evolutionary stages of the YSOs.} 
   {The region was imaged at 3, 6, and 20 cm using 
the Australia Telescope Compact Array. Fluxes and spectral indices 
for the brightest radio sources were derived from these
observations. Some of the 3 and 6 cm data were obtained simultaneously,
which is useful for reliable spectral index determinations of
variable sources.}
   {Eight pointlike sources were detected. Seven of them can be 
assigned to YSOs, which have counterparts in the X-rays, infrared or 
submm. One of the YSOs, Radio Source 9, is a Class 0 candidate, and 
another, IRS 7B, is suggested to be in the Class 0/I transition 
stage. IRS 7B is associated with extended radio lobes at 6 and 20 
cm. The lobes may have a gyrosynchrotron emission component, which 
could be understood in terms of Fermi accleration in shocks. The Class 
I objects detected here seem to be a mixed lot. One of these, the wide 
binary IRS 5, shows a negative spectral index, rapid variability, and a 
high degree of circular polarisation with $V/I\approx33$ \% on one of the 
days of observation. These signs of magnetic activity suggest that at 
least one of the binary components has advanced beyond the Class I 
stage. The radio source without YSO assigment, Radio Source 5, has 
been suggested to be a brown dwarf. The radio properties, in particular 
its persistent strong emission, do not support this classification.} 
   {The radio characteristics of the detected YSOs roughly agree with the
scheme where the dominant emission mechanism changes with age.
The heterogeneity of the Class I group can possibly be explained by
a drastic decline in the jet activity during this stage, which also changes
the efficiency of free-free absorption around the protostar.} 
 
   \keywords{stars: formation -- ISM: clouds -- ISM: individual objects: 
R CrA cloud -- ISM: jets and outflows -- radio continuum: ISM} 
 
   \maketitle 
% 
%________________________________________________________________ 
 
\section{Introduction} 
 
The presence of a protostar in a molecular cloud core is indicated 
by compact centimetre radio emission, a bipolar molecular outflow, and/or near-
or mid-infrared (NIR/MIR) emission (\cite{andre2000}).
Centimetre continuum radiation observed toward YSOs
originates in most cases from thermal free-free emission from
partially ionized stellar winds or jets and/or accretion surfaces.
Non-thermal gyrosynchrotron radiation has been observed towards
the YSO IRS 5 in the R CrA cloud (\cite{feigelson1998}).
Synchrotron emission has been identified via detection of linear
polarisation towards a very young T Tauri star by Phillips et al. (1996)
(see also \cite{massi2006}). 
There is also evidence for extended synchrotron protostellar jets 
presented in the triple radio source in Serpens
(\cite{rodriguez1989}).

The Corona Australis (CrA) dark cloud is a nearby ($d \sim 170$ pc,
\cite{knude1998})\footnote{There are also shorter distance estimates
for the CrA star-forming region. Marraco \& Rydgren (1981)
derived a distance of only 129 pc for the R CrA dark cloud.
Also, Casey et al. (1998) derived a similar distance of $129\pm11$ pc
for TY CrA. In the present paper, the Hipparcos distance $\sim 170$ pc
is adopted.} low- and intermediate-mass star-forming region,
where the R CrA core represent the most active site of star formation.
The evolutionary stages range from  
Class 0 objects to pre-main sequence (PMS) stars (\cite{wilking1997};
\cite{nisini2005}; \cite{nutter2005}).
The R CrA region is therefore favourable for studies of the dependence of
radio continuum emission on the stellar evolutionary stage.

In recent years, the cluster of YSOs in the R CrA core, also called 
the \textit{Coronet} (\cite{taylor1984}), have been studied extensively in 
X-rays, in the infrared, submillimetre, and in radio (\cite{hamaguchi2005};  
\cite{nutter2005}; \cite{forbrich2006}; 2007; \cite{forbrich2007b}; 
\cite{groppi2007}). 
It has become evident that some of the radio and X-ray sources in the 
region are highly variable. 
Forbrich et al. (2007) performed the first simultaneous X-ray, radio, NIR, 
and optical monitoring of the cluster, and detected seven objects 
simultaneously in all wavelength bands. These series of studies have given 
information on the processes and emission 
mechanisms associated with various stages of early stellar evolution, 
and helped to ascertain the evolutionary stages of most of the cluster 
members. However, the Radio Sources 5 and 9, first detected by Brown (1987), 
have not yet been unambiguously classified. These two sources will
be hereafter called as Brown 5 and Brown 9.

In this paper we present centimetre continuum imaging of the R CrA 
core performed at the Australia Telescope Compact Array using 
the wavelengths 3, 6 and 20 cm. 
The purpose of this study was to determine the radio spectral 
indices of the various sources in the region, and thereby remove some 
uncertainty concerning their nature. At the same we add a couple of 
measurement points to the time series of the varying sources. 
In Sect. 2 of this paper we review briefly the radio continuum 
emission mechanisms associated with YSOs.
The observations and data reduction procedures are described in 
Sect. 3. The results are presented in Sect. 4. In Sect. 5 we discuss the 
results, and in Sect. 6 we summarise our conclusions. 

\section{Radio continuum emission from YSOs}

\subsection{Spectral index and polarisation}

Radio emission from young protostars usually has a spectral 
index\footnote{The spectral index is defined as 
$S_{\nu} \propto \nu^{\alpha}$, where $S$ is the flux 
density and $\nu$ is the frequency.}, $\alpha$,  
in the range [-0.1, 2.0]. This range is characteristic for  
thermal free-free emission from partially ionized gas 
($n_{\rm H^+}/n_{\rm H}\sim0.1$, see \cite{martin1996};  
\cite{anglada1998} and references therein). 
The $\nu^2$ dependence represents optically thick 
free-free emission, whereas optically thin emission is almost
independent of frequency (e.g., \cite{gudel2002}).  
Conforming with this emission mechanism the brightness temperatures measured 
for protostellar sources are moderate ($T_{\rm b} \leq 10^4$ K). 
The emission can originate in ionized jets from protostars or in shocks 
produced by jets penetrating into the circumstellar material 
(e.g., \cite{curiel1987}). 
Other possible sources are ionized disk wind (e.g., \cite{martin1996}) 
or accretion shocks on circumstellar disks (e.g., \cite{winkler1980}; 
\cite{neufeld1996}; see also \cite{ghavamian1998}). 

The predicted spectral index of the so-called standard spherical 
stellar wind (i.e., fully ionized and isotropic stellar wind with constant 
velocity) is 0.6 at cm-wavelengths (e.g., \cite{panagia1975}). 
According to the model of Bertout (1983) also accretion
can produce a spectral index of 0.6. 
In this model the region emitting at radio wavelengths is photoionized by 
soft X-ray and EUV radiation from an accretion shock around 
a protostellar core. In contrast, in the model of Felli et al. (1982) 
accreting ionized circumstellar envelope produces a flat spectrum with
spectral indices in the range of $\alpha=-0.1 - 0.1$. 

Reynolds (1986) has modelled continuum emission from both collimated and  
accelerated spherical flows. According to these results, $\alpha$ can
increase above 0.6 for an accelerated flow, but a collimated ionized flow
produces a flat spectrum with $\alpha < 0.6$. In the model of collimated  
ionized flow, the opacity decreases with the distance to the  
central source (see also \cite{anglada1998}). 

Several YSOs show negative spectral indices, $\alpha<-0.1$, in the radio, 
characteristic of non-thermal gyrosynchrotron emission (see, e.g., 
\cite{forbrich2006}; \cite{stamatellos2007} and references therein). 
Gyrosynchrotron emission arises from mildly relativistic 
electrons (Lorentz factor $\gamma \lesssim 2-3$, see \cite{dulk1985})
gyrating in magnetic fields close to the YSO and/or to the star-disk
interaction region. Thus, non-thermal radio emission indicates
magnetic activity around protostars. 

Also shocks can accelerate electrons to mildly relativistic
energies ($\sim 1$ MeV). Clear evidence of efficient electron
acceleration associated with YSOs have been found in the Serpens triple
source (see \cite{feigelson1999}).  
Gyrosynchrotron radiation can be generated in jet induced 
shocks by Fermi acceleration of electrons. 
In the so-called diffusive shock acceleration (DSA)  
electron scatters off magnetic irregularities frozen in to the local plasma 
flow. This interaction can change the direction of electron, which can result 
in several encounters with the shock front enabling several encounters with 
the shock. The shock crossings lead to a systematic gain of kinetic energy. 
Besides having a negative spectral index, gyrosynchrotron emission is 
characterised by rapid variability, moderate degree of circular 
polarisation ($|V|/I\lesssim20$ \%), and very high 
brightness temperature ($T_{\rm b} \geq 10^{7.5}$ K) 
(see, e.g., \cite{andre1996}; \cite{forbrich2006}). 

Partial linear polarisation characteristic of synchrotron radiation  
from ultrarelativistic electrons ($\gamma \gg 1$) was detected by  
Phillips et al. (1996) (see Sect. 1), which implies very efficient electron 
acceleration. As a distinction from the gyrosynchrotron radiation, 
the degree of circular polarisation (which to order of magnitude amounts to 
$\approx \gamma^{-1}$) of synchrotron radiation is low (\cite{dulk1985}). 
 
The observed spectral index mainly describes the source average 
opacity and in the case of inhomogeneous source, the radio emission  
can originate from both optically thin and thick regions.   
Thus, the spectral index alone gives only a rough diagnostic of the nature 
of YSO radio emission (\cite{andre1996}; \cite{anglada1998}). 

\subsection{Relation between radio continuum emission mechanism and 
evolutionary stage of the YSO} 

The dominant radio continuum emission mechanism is suggested to change  
with the age of the YSO (\cite{gibb1999}). 
In the Class 0/I stage, the principal 
emission mechanism is thermal free-free emission 
(accretion and outflowing jets). Continuum emission from Class II objects 
(classical T Tauri stars, CTTSs) at cm-wavelengths is caused by 
thermal free-free mechanism in strong ionized winds. 
In the Class I/II phase, possible non-thermal component is likely to 
be free-free absorbed in the surrounding dense ionized gas. The detection 
of a non-thermal component becomes more and more difficult towards higher 
frequencies where free-free absorption is accentuated (\cite{gudel2002}). 
Later, in the Class III stage (weak-lined T Tauri stars, WTTSs), 
when circumstellar disks and ejecta are very weak or absent, 
non-thermal gyrosynchrotron emission from the exposed 
star can dominate at radio wavelengths due to increased stellar magnetic
field (see \cite{feigelson1999}; \cite{gibb1999}; \cite{wilking2001}). 

Free-free emission can dominate gyrosynchrotron emission not only if  
the density is high (free-free absorption), but also if the temperature  
is low or the magnetic field is weak. Moreover, different frequency depencies 
may cause switching between the dominant emission mechanism at a certain 
frequency (\cite{dulk1985}). Furthermore, Gibb (1999) suggested that
the apparent correlation between radio continuum emission and evolutionary
stage could be due to source geometry and optical depth effects.

\section{Observations and data reduction} 

The R CrA star-forming region was observed at 3, 6 and 20 cm continuum
with the Australia Telescope Compact Array (ATCA)\footnote{Operated by
the CSIRO Australia Telescope National Facility.},
located near Narrabri, Australia. 
The observations were made during two observing runs in 1998 and 2000. 
The dates, configurations, phase centres, calibrators, and other observing 
parameters are given in Table~\ref{table:1}. The correlator configuration used 
provided at each frequency a total bandwith of 128 MHz recorded as a 16 
channel spectrum. Four products of the linearly polarised feeds were recorded 
(XX, YY, XY and YX), and thus all the Stokes parameters can be 
solved from the calibrated data. 
 
The data were calibrated and imaged using the Miriad software package 
(\cite{sault1995}). The visibilities were transformed into maps using natural 
weighting. The image deconvolution was done with
the Miriad task {\tt mfclean}. 
Primary beam correction was finally done for images
(Miriad task {\tt linmos}). Intensity maxima exceeding 3 times
the local rms are considered as detections.
For non-detections we give $3\sigma$ upper limits. 
 
% Table 1 
\begin{table*} 
\caption{Observational parameters.} 
\begin{minipage}{2\columnwidth} 
\centering 
\renewcommand{\footnoterule}{} 
\label{table:1} 
\scriptsize 
\begin{tabular}{c c c c c c} 
\hline\hline  
Date of observations & 1998 Jan 9/10 & 2000 May 28 & 2000 May 31 & 2000 Jun 9 & 2000 Jun 10\\ 
Time (UT) interval & 19:06:05-08:40:55 & 09:56:35-23:11:45 & 09:33:35-21:56:25 & 09:26:15-22:21:55 & 09:25:55-22:03:35\\ 
Wavelength & 3 and 6 cm & 3 cm & 3 cm & 6 cm & 20 cm\\  
Bandwith & 128 ($16\times8$) MHz\footnote{Used in all cases.}\\ 
Configuration & 1.5A & 1.5A & 1.5A & 6B & 6B\\ 
$uv$-distance range & 5--149 and 2.6--74.5 k$\lambda$ & 5--149 k$\lambda$ & 5--149 k$\lambda$ & 3.6--99.5 k$\lambda$ & 1--30 k$\lambda$\\ 
Phase centre & $19^{\rm h}01^{\rm min}44\fs4$ & $19^{\rm h}01^{\rm min}56\fs5$ & $19^{\rm h}01^{\rm min}56\fs5$ & $19^{\rm h}01^{\rm min}56\fs5$ & $19^{\rm h}01^{\rm min}56\fs5$\\ 
($\alpha_{2000.0}$, $\delta_{2000.0}$) & $-36\degr58\arcmin40\farcs3$ & $-36\degr57\arcmin28\arcsec$ & $-36\degr57\arcmin28\arcsec$ & $-36\degr57\arcmin28\arcsec$ & $-36\degr57\arcmin28\arcsec$\\ 
Flux calibrator\footnote{Calibrator flux densities are the following: PKS 1934-638: $S_{3 \, \rm cm}=2.84$ Jy, $S_{6 \, \rm cm}=5.83$ Jy, $S_{20 \, \rm cm}=14.95$ Jy; B1849-36: $S_{3 \, \rm cm}=0.42$ Jy, $S_{6 \, \rm cm}=0.70$ Jy; B1933-400: $S_{3 \, \rm cm}=2.09$ Jy, $S_{6 \, \rm cm}=1.38$ Jy, $S_{20 \, \rm cm}=0.81$ Jy (from the ATNF web site {\tt http://www.narrabri.atnf.csiro.au/calibrators/}).} & PKS 1934-638 & PKS 1934-638 & PKS 1934-638 & PKS 1934-638 & PKS 1934-638 \\ 
Phase calibrators$^b$ & B1849-36 & B1933-400 & B1933-400 & B1933-400 & B1933-400\\
Image processing & Naturally weighted$^a$\\ 
FWHM of synthesized beam & $3\farcs4 \times 3\farcs1$ and $6\farcs5 \times 5\farcs7$ & $4\farcs4 \times 3\farcs3$ & $4\farcs4 \times 3\farcs2$ & $4\farcs2 \times 2\farcs3$ & $13\farcs1 \times 6\farcs6$\\ 
Beam position angle & $-31\fdg0$ and $-19\fdg6$\footnote{The difference between the beam position angles is due to the different amount of flagged $uv$ data.} & $2\fdg5$ & $1\fdg9$ & $-2\fdg8$ & $0\fdg1$\\ 
rms noise level [$\mu$Jy beam$^{-1}$] & 97 and 81 & 76 & 47 & 44 & 120\\ 
\hline  
\end{tabular}  
\end{minipage} 
\end{table*}

% Table 2 
\begin{table*} 
\caption{Positions, identification and the class of the detected 
radio sources.} 
\begin{minipage}{2\columnwidth} 
\centering 
\renewcommand{\footnoterule}{} 
\label{table:2} 
\begin{tabular}{c c c c c} 
\hline\hline  
& \multicolumn{2}{c}{Peak position\footnote{These positions refers to peak positions at 3 cm observed in January 1998.}} & \\ 
No. & $\alpha_{2000.0}$ [h:m:s] & $\delta_{2000.0}$ [${\degr}$:${\arcmin}$:${\arcsec}$] & Source ID & Source class\\ 
\hline 
1......& 19 01 43.3 & -36 59 13 & Brown 5\footnote{Source number in \cite{brown1987}.} & unclear\\ 
2......& 19 01 48.0 & -36 57 22 & IRS 5 & Class I\\ 
3......& 19 01 50.5 & -36 56 38 & IRS 6 & Class II (CTTS)\\ 
4......& 19 01 50.7 & -36 58 10 & IRS 1 & Class I\\ 
5......& 19 01 53.6 & -36 57 12 & R CrA & HAe\\ 
6......& 19 01 55.2 & -36 57 17 & Brown 9$^b$ & Class 0 ?\\ 
7......& 19 01 55.3 & -36 57 22 & IRS 7A, IRS 7W\footnote{Source used in \cite{forbrich2006}; 2007.} & Class I\\ 
8......& 19 01 56.4 & -36 57 28 & IRS 7B, IRS 7E$^c$ & Class 0/I\\ 
\hline  
\end{tabular}  
\end{minipage} 
\end{table*}

% Table 3 
\begin{table*} 
\caption{Flux densities with $\pm 1\sigma$ errors of the detected 
radio sources at 3, 6 and 20 cm. The values are corrected for the primary 
beam response.} 
\begin{minipage}{2\columnwidth} 
\centering 
\renewcommand{\footnoterule}{} 
\label{table:3} 
\begin{tabular}{c c c c c c c} 
\hline\hline 
 Source & \multicolumn{2}{c}{1998 Jan 9/10} & 2000 May 28 & 2000 May 31 & 2000 Jun 9 & 2000 Jun 10\\ 
       & $S_{3 \, \rm cm}$ [mJy] & $S_{6 \, \rm cm}$ [mJy] & $S_{3 \, \rm cm}$ [mJy] & $S_{3 \, \rm cm}$ [mJy] & $S_{6 \, \rm cm}$ [mJy] & $S_{20 \, \rm cm}$ [mJy]\\ 
\hline 
Brown 5 & $0.96\pm0.06$ & $1.42\pm0.10$ & $1.40\pm0.13$ & $1.49\pm0.20$ & $1.05\pm0.01$ & $1.80\pm0.02$\\ 
IRS 5 & $0.81\pm0.03$ & $1.06\pm0.04$ & $1.87\pm0.21$ & $0.70\pm0.04$ & $0.62\pm0.03$ & $0.40\pm0.04$\\ 
IRS 6 & - & $0.46\pm0.04$ & - & - & - & -\\ 
IRS 1 & $0.33\pm0.04$ & - & $0.62\pm0.21$ & $0.52\pm0.06$ & $0.21\pm0.02$ & -\\ 
R CrA & $0.39\pm0.03$ & - & $0.46\pm0.02$ & $0.35\pm0.04$ & $0.29\pm0.03$ & -\\ 
Brown 9\footnote{Blended with IRS 7A at 6 cm in the dataset 1998 Jan 9/10, and at 20 cm in 
the dataset 2000 Jun 10.} & $0.74\pm0.14$ & - & $1.88\pm0.08$ & $1.93\pm0.10$ & $1.58\pm0.11$ & -\\ 
IRS 7A\footnote{For these extended sources we give the integrated
flux densities over the pixels above the $3\sigma$ level as calculated by the Miriad task {\tt imfit}.} & $4.19\pm0.07$ & $5.49\pm0.12$ & $6.00\pm0.02$ & $5.90\pm0.01$ & $5.60\pm0.09$ & $5.10\pm0.08$\\ 
IRS 7B$^b$ & $2.56\pm0.06$ & $2.75\pm0.03$ & $3.65\pm0.07$ & $3.70\pm0.07$ & $3.42\pm0.06$ & $3.67\pm0.06$\\ 
\hline  
\end{tabular}  
\end{minipage} 
\end{table*}

% Table 4 
\begin{table*} 
\caption{Spectral indices with $\pm 1\sigma$ errors of the observed 
sources.} 
\begin{minipage}{2\columnwidth} 
\centering 
\renewcommand{\footnoterule}{} 
\label{table:4} 
\begin{tabular}{c c c c} 
\hline\hline 
Source & $\alpha_{3 \, \rm cm}^{6 \, \rm cm}$ \footnote{$\alpha_{3 \, \rm cm}^{6 \, \rm cm}$ derived from 1998 Jan 9/10 data  
using convolved flux densities.} & $\alpha_{3 \, \rm cm}^{6 \, \rm cm}$ \footnote{$\alpha_{3 \, \rm cm}^{6 \, \rm cm}$ derived from 2000 May 31 \& Jun 9 
data using original flux densities.} & $\alpha_{6 \, \rm cm}^{20 \, \rm cm}$ \footnote{$\alpha_{6 \, \rm cm}^{20 \, \rm cm}$ derived from 2000 Jun 9/10 data  
using convolved flux densities.}\\ 
\hline 
Brown 5 & $-0.34\pm0.12$ & $0.76\pm0.29$ & $-0.30\pm0.01$\\ 
IRS 5 & $-0.12\pm0.07$ & $0.26\pm0.16$ & $0.54\pm0.08$\\ 
IRS 1 & - & $1.96\pm0.32$ & -\\ 
R CrA & - & $0.41\pm0.33$ & -\\ 
Brown 9 & - & $0.43\pm0.19$ & -\\ 
IRS 7A & $0.19\pm0.04$ & $0.11\pm0.03$ & $0.34\pm0.02$\\ 
IRS 7B & $0.38\pm0.02$ & $0.17\pm0.06$ & $-0.05\pm0.01$\\ 
\hline  
\end{tabular}  
\end{minipage} 
\end{table*}

\section{Results} 

\begin{figure}[!h] 
\resizebox{\hsize}{!}{\includegraphics[angle=-90]{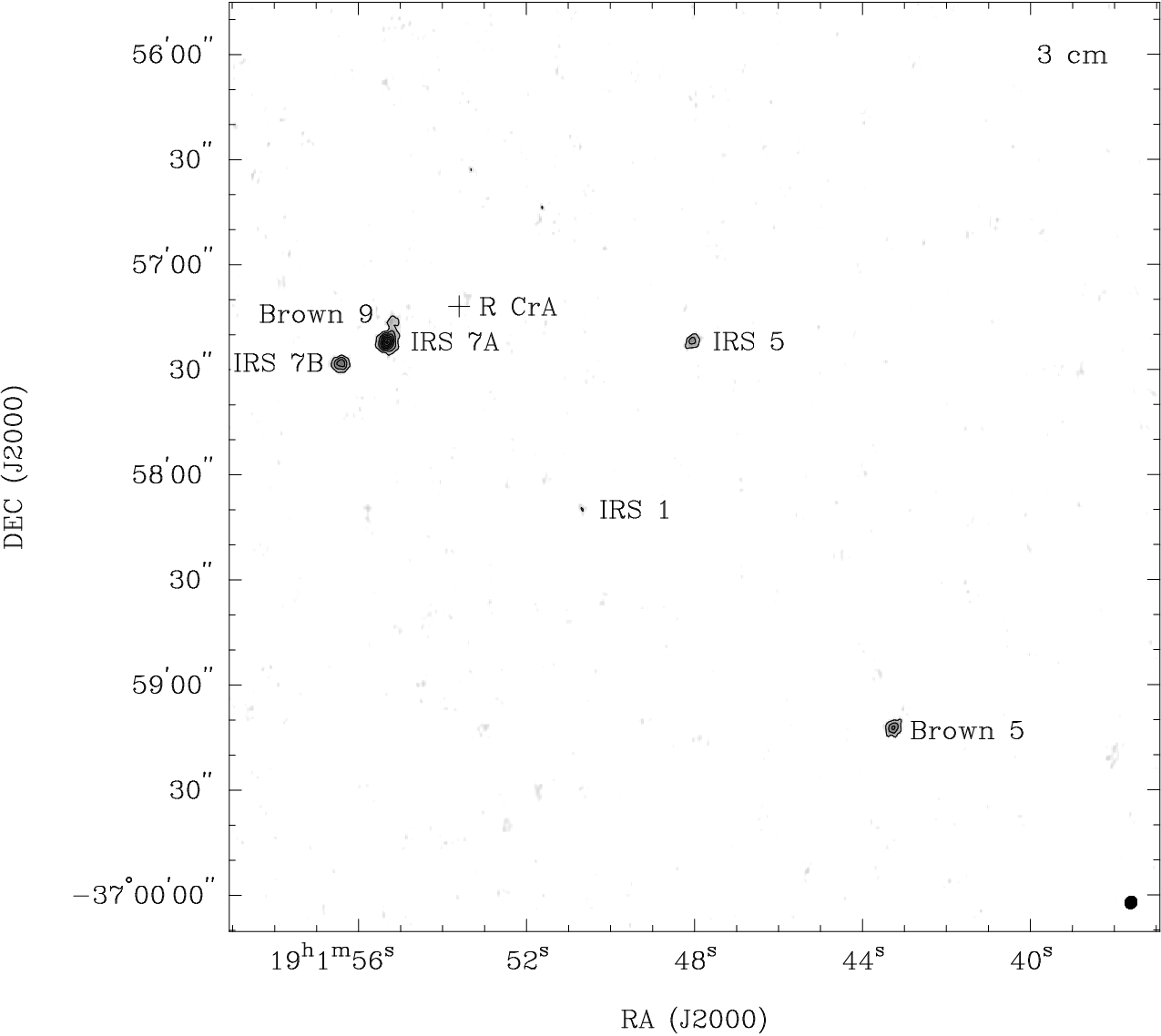}} 
\caption{Map of the 3 cm continuum made from the 1998 Jan 9/10 visibility data 
toward the R CrA star-forming region. The logarithmic grey-scale
range from 0.15 to 5 mJy beam$^{-1}$. The contours are from
0.29 mJy beam$^{-1}$ ($3\sigma$) to 2.9 mJy beam$^{-1}$ in steps of
0.29 mJy beam$^{-1}$.
Shown at the bottom right-hand corner is the synthesized beam:
FWHM $=3\farcs4 \times 3\farcs1$ and P.A.$=-31\fdg0$.} 
\label{figure1} 
\end{figure}

The 3 and 6 cm continuum images from 1998 Jan 9/10 are shown in
Figs.~\ref{figure1} and \ref{figure2}, respectively. 
The 3 and 6 cm images from 2000 May 31 and Jun 9 toward the
IRS 7 region are shown in Fig.~\ref{figure3}. 
A comparison of the 20 and 6 cm images from 2000 Jun 10 and 1998
Jan 9/10 showing the radio lobes associated with IRS 7B are presented in
Fig.~\ref{figure4}.
Altogether 8 radio sources were identified by 
inspecting the maps visually. The sources are listed in Table~\ref{table:2}.
This gives their J2000 coordinates, identifications with 
previously known sources, and suggested classifications.

The flux densities are given in Table~\ref{table:3}. These were
obtained from Gaussian fits using the Miriad task {\tt imfit}.
For the pointlike sources 1 -- 6 the flux density 
is the peak of the Gaussian, whereas for the somewhat extended 
sources 7 and 8 we give the integrated intensity over the pixels above  
the $3\sigma$ level as calculated by {\tt imfit}. 
The errors are determined in the fitting procedure. 
As can be seen from Table~\ref{table:3}, only Brown 5, IRS 5, and 
the IRS 7A/B pair are visible at all three wavelengths. 

We also obtained spectral indices between 3 and 6 cm 
($\alpha_{3 \, \rm cm}^{6 \, \rm cm}$),  and 6 and 20 cm 
($\alpha_{6 \, \rm cm}^{20 \, \rm cm}$). 
For the determination of these spectral indices, the maps at the two
frequencies were convolved to the same angular resolution using the Miriad
task {\tt convol}. On May 31 and Jun 9, 2000 we used configurations
which gave similar synthesized beams at 3 and 6 cm in order to derive reliable 
spectral indices for extended sources.
In 1998, Jan 9/10, the 3 and 6 cm observations were done simultaneously,
and in this case the spectral indices should be more reliable for variable
point sources.
The spectral indices $\alpha_{3 \, \rm cm}^{6 \, \rm cm}$ and
$\alpha_{6 \, \rm cm}^{20 \, \rm cm}$, and their $1\sigma$ errors
are given in Table~\ref{table:4}. The error estimates are based on the rms
errors of the flux densities, and do not take into account the 
possible errors caused by different beams or source variability.

\begin{figure}[!h] 
\resizebox{\hsize}{!}{\includegraphics[angle=-90]{6cm.ps}} 
\caption{Map of the 6 cm continuum made from the 1998 Jan 9/10 
visibility data toward the R CrA star-forming region.  
The logarithmic grey-scale range from 0.1 to 5 mJy beam$^{-1}$.
The contours are from 0.24 mJy beam$^{-1}$ ($3\sigma$) to 2.4 mJy beam$^{-1}$
in steps of 0.24 mJy beam$^{-1}$. Shown at the bottom right-hand corner
is the synthesized beam: FWHM $=6\farcs5 \times 5\farcs7$ and
P.A.$=-19\fdg6$.} 
\label{figure2} 
\end{figure} 
 
\begin{figure}[!h]
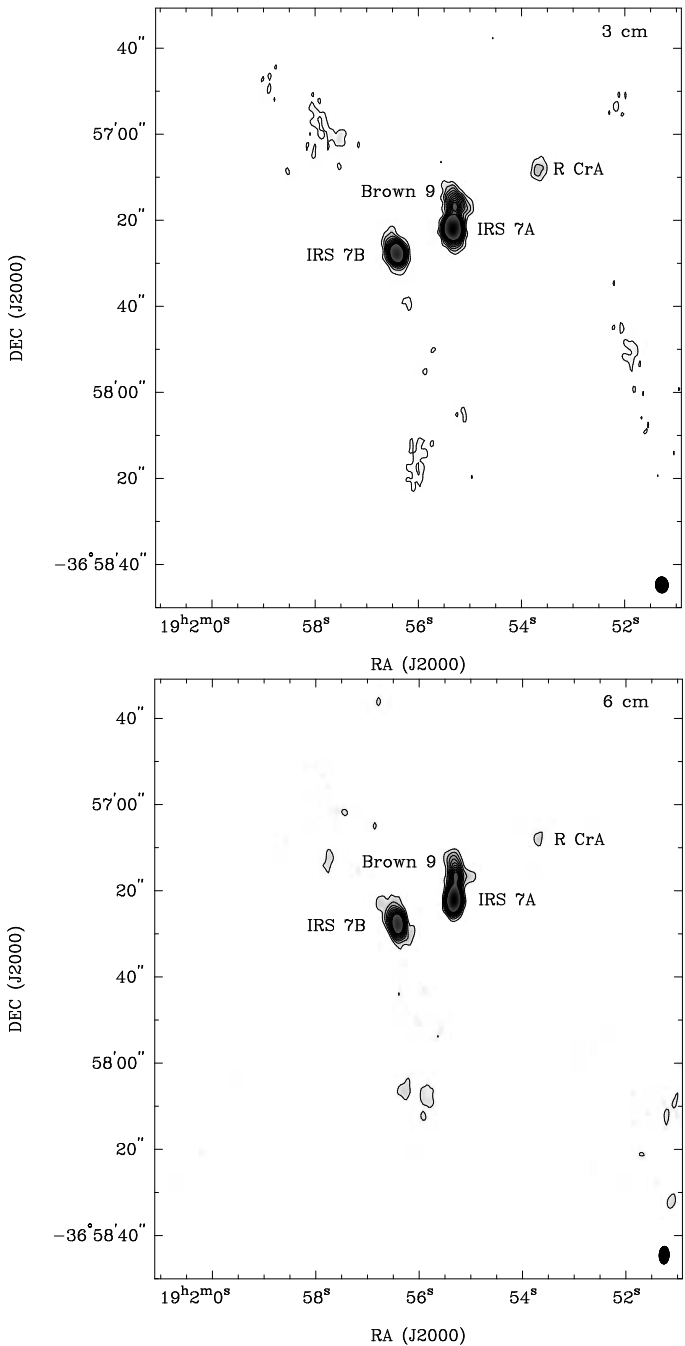
 
\begin{center} 
\resizebox{\hsize}{!}{\includegraphics[angle=-90]{3cm_310500.ps}}  
\resizebox{\hsize}{!}{\includegraphics[angle=-90]{6cm_090600.ps}}  
\caption{\textbf{Top}: The 3 cm contour map made from the 2000 
May 31 visibility data toward the IRS 7 region. 
The contours are from 0.14 mJy beam$^{-1}$ 
($3\sigma$) to 1.4 mJy beam$^{-1}$ in steps of 0.14 mJy beam$^{-1}$. 
The logarithmic grey-scale range from 0.15 to 5 mJy beam$^{-1}$.
Shown at the bottom right-hand corner is the synthesized beam: 
FWHM $=4\farcs4 \times 3\farcs2$ and P.A.$=1\fdg9$. \textbf{Bottom}: 
The 6 cm contour map made from the 2000 Jun 9 visibility data toward 
the IRS 7 region. The contours are from 0.13 mJy beam$^{-1}$ ($3\sigma$)
to 1.3 mJy beam$^{-1}$ in steps of 0.13 mJy beam$^{-1}$. 
The logarithmic grey-scale range from 0.1 to 5 mJy beam$^{-1}$.
Shown at the bottom right-hand corner is the synthesized beam:
FWHM $=4\farcs2 \times 2\farcs3$ and P.A.$=-2\fdg8$.} 
\label{figure3} 
\end{center} 
\end{figure} 
 
\begin{figure}[!h] 
\resizebox{\hsize}{!}{\includegraphics[angle=-90]{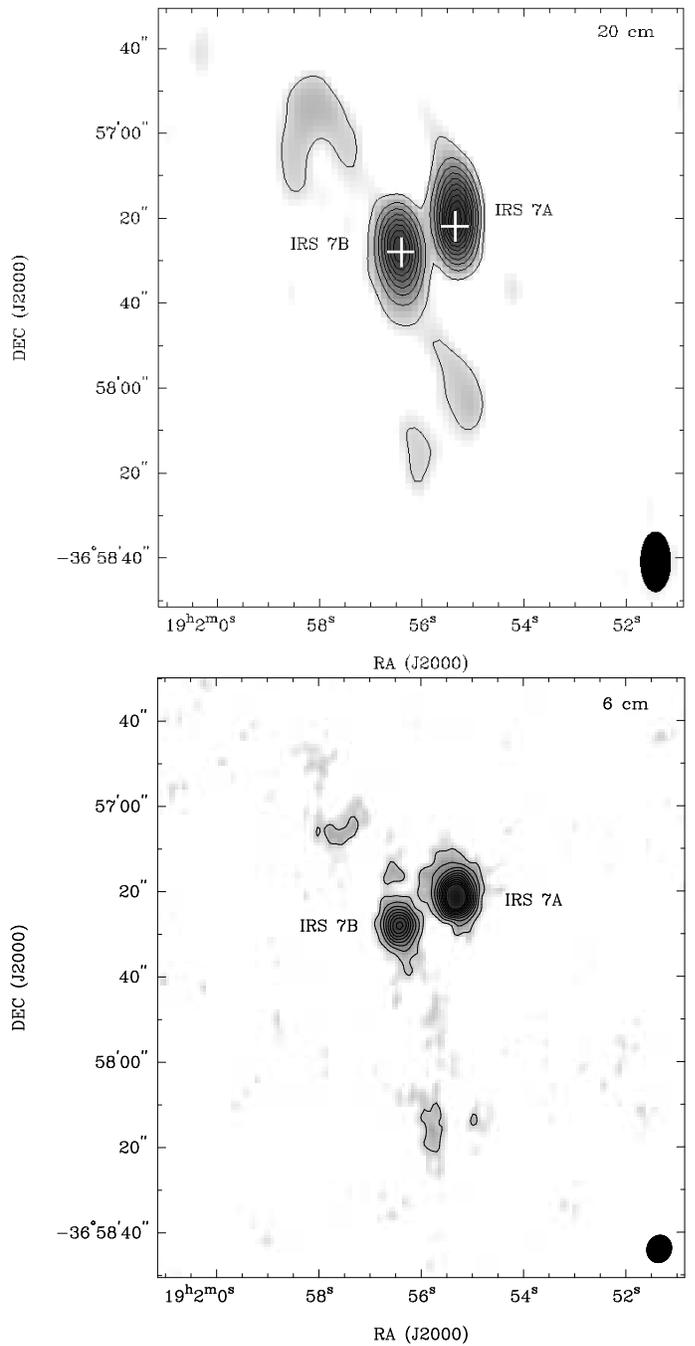}} 
\resizebox{\hsize}{!}{\includegraphics[angle=-90]{6cm_090198.ps}} 
\caption{\textbf{Top}: The 20 cm contour map made from the 2000 Jun 
10 visibility data toward the IRS 7 region. The contours are from 0.36 mJy 
beam$^{-1}$ ($3\sigma$) to 3.6 mJy beam$^{-1}$ in steps of 0.36 
5mJy beam$^{-1}$. The logarithmic grey-scale range from 0.23
to 7 mJy beam$^{-1}$.
White crosses shows the 3 and 6 cm peak positions. 
Shown at the bottom right-hand corner is the synthesized beam: 
FWHM $=13\farcs1 \times 6\farcs6$ and P.A.$=0\fdg1$. \textbf{Bottom}: 
Blow-up of Fig.~\ref{figure2} showing the IRS 7 region.
Shown at the bottom right-hand corner is the synthesized beam: 
FWHM $=6\farcs5 \times 5\farcs7$ and P.A.$=-19\fdg6$.} 
\label{figure4} 
\end{figure} 
 
\section{Discussion} 
 
All the eight sources detected in the present study are known from 
previous radio surveys (\cite{brown1987} (2 and 6 cm, VLA);  
\cite{suters1996} (2, 3.5, 6, and 20 cm, VLA and ATCA); 
\cite{feigelson1998} (3.5 cm, VLA); \cite{choi2004} (0.7 cm, VLA); 
\cite{forbrich2006}; 2007 (3.6 cm, VLA)). 
Five of the sources, IRS 5, IRS 6, IRS 1, R CrA, and IRS 7A, have 
counterparts both in the near/mid-infrared and X-rays. 
Moreover, IRS 7B, has been recently detected in the mid-infrared
(but not in the NIR) and X-rays. Brown 9 corresponds to
a submm peak (\cite{groppi2007}).
Brown 5 has only been detected in radio, but it 
may be associated with a near-IR source (\cite{wilking1997}). In what follows 
we discuss the radio properties of the detected sources in detail. 
 
\subsection{Comments on individual sources} 
 
\subsubsection{The IRS 7 region} 
 
IRS 7 is the most active complex in the \textit{Coronet} cluster. 
A dense, bipolar molecular outflow and a large rotating molecular 
disk (outer radius $>3900$ AU at $d=170$ pc) were detected in the IRS 7 region 
(\cite{anderson1997a}; 1997b). Groppi et al. (2004) found that the envelope 
surrounding the disk show signs of infall in the HCO$^+(4-3)$ spectra. 
The region contains three compact radio sources, IRS 7A, IRS 7B,
and Brown 9. The rotation axis of the molecular envelope coincides with
a string of HH objects (\cite{hartigan1987}; see Fig.~10 in
\cite{anderson1997a}). These features suggest ongoing powerful accretion
and possible shocks associated with infall onto an accretion disk.

IRS 7A and IRS 7B are the brightest radio sources in the whole cloud.
The evolutionary stages of these sources have been
discussed by Wilking et al. (1997), Choi \& Tatematsu (2004), Hamaguchi et al. 
(2005), Forbrich et al. (2006), and Groppi et al. (2007). 
Brown 9 (\cite{brown1987}) is located $5\arcsec$ north of IRS 7A 
(850 AU at $d=170$ pc). Brown 9 and IRS 7A are blended in the 6 cm map from 
1998 and in the 20 cm map from 2000. 
 
IRS 7A has not shown much radio variability. The flux 
densities of IRS 7A are very similar to those observed by Brown (1987), 
Feigelson et al. (1998) and Forbrich et al. (2006).  
The flat spectrum with $\alpha_{3 \, \rm cm}^{6 \, \rm cm}=0.19\pm0.04$ and 
$0.11\pm0.03$ and $\alpha_{6 \, \rm cm}^{20 \, \rm cm}=0.34\pm0.02$
can be explained by optically thin emission from ionized
circumstellar envelope or a collimated jet. 
IRS 7A has been detected both in the near-infrared and in the X-ray band. 
In view of the relatively bright radio emission, these characteristics 
suggest that the source represents the protostellar Class I.  

The relationship between IRS 7A, Brown 9 and
the nearby 10$\mu$m source between the two (\cite{wilking1997})
has been discussed in Choi \& Tatematsu (2004, hereafter CT).
They suggest that these sources belong to a system formed by a central star
(close to the 10$\mu$m source, associated with the $\lambda=6.9$ mm source
No. 3 of CT) and ionized jets (IRS 7A and Brown 9) on its both sides.
By inspecting the map of CT, the $\lambda=3.6$ cm map of
Forbrich et al. (2006), and the SCUBA 450 $\mu$m continuum map of
Groppi et al. (2007), one agrees that the three sources can be physically
connected. It is, however, unlikely that the X-ray source IRS 7A and the submm 
source Brown 9 would just be jets arising from {the 10$\mu$m source associated
to IRS 7}.
In all other near-IR sources of our sample the hottest spot, i.e. the X-ray 
source, shows the location of the central protostar. 
On the other hand, the submm/far-IR spectrum of Brown 9 is not
possible to explain with free-free emission from plasma, but it is
characteristic of a Class 0 protostar (see Fig.~8 in \cite{groppi2007}).

The flux densities of Brown 9 at 3 cm found in the present study  
are mostly similar to those found in the previous studies 
(\cite{feigelson1998}; \cite{forbrich2006}). However, the 
source shows a significant brightening (by a factor of two) since the 
observation of Brown (1987). The spectral index,  
$\alpha_{3 \, \rm cm}^{6 \, \rm cm}=0.43\pm0.19$, is consistent with 
thermal free-free emission originating in a collimated ionized flow. 
The source of this flow is likely to be embedded in core detected 
with SCUBA.  

The IRS 7B flux densities derived here are  
$\sim$0.5--1.5 mJy higher than those reported 
by Brown (1987), Feigelson et al. (1998) and Forbrich et al. (2006).  
This difference can be attributed to source variablity. The spectral indices 
between 3 and 6 cm are $\alpha_{3 \, \rm cm}^{6 \, \rm cm}=0.38\pm0.02$ and 
$0.17\pm0.06$. The spectral index between 6 and 20 cm, 
$\alpha_{6 \, \rm cm}^{20 \, \rm cm}=-0.05\pm0.01$, is close to 
the limit of thermal emission and non-thermal gyrosynchrotron 
emission ($\alpha=-0.1$). 
We note, however, that $\alpha_{3 \, \rm cm}^{6 \, \rm cm}$ is derived
from simultaneous observations (1998, Jan 9/10) and is much more significant
than the value for $\alpha_{6 \, \rm cm}^{20 \, \rm cm}$ ($5\sigma$) based on 
observations from two consecutive days.

IRS 7B is associated with weak, extended radio emission. The lobes on
the northeastern and southern sides of the point source are clearly
visible on the 6 cm map from 1998 and the 20 cm map from 2000 (see
Fig.~\ref{figure4}). These two maps were obtained using configurations
with the shortest $uv$-distances (see Table~\ref{table:1}).
Traces of these lobes can be seen in the 3 and 6 cm images from 2000
(Fig.~\ref{figure3}), which represent the highest angular resolution available
here. The lobes become visible also in these data by applying a Gaussian taper
which produces a resolution similar to that in the 20 cm image.
The S/N ratio is, however, too poor to compare individual features at 3 and
6 cm. The extended emission was already detected by Brown (1987) at 6 cm
and by Suters et al. (1996) at 20 cm. The radio lobes may be associated
with a stage prior to the formation of a HH object (\cite{curiel1993}). 
The fact that the lobes are more prominent at 6 and 20 cm than at 3 cm
or shorter wavelengths suggests a negative spectral index and consequenly,
a possible gyrosynchrotron emission component. 
This led Harju et al. (2001) to suggest that the source could be a distant
radio galaxy or a Galactic microquasar.  
Later observations have, however, ruled out this possibility. 
Pontoppidan et al. (2003) detected IRS 7B in the 4.67 $\mu$m 
absorption lines of solid CO, which clearly originate in circumstellar 
material. Moreover, Hamaguchi et al. (2005) pointed out that 
it would be very unlike to observe AGN with X-ray flux of IRS 7B in the 
IRS 7 region. Also the X-ray spectra of IRS 7B does not resemble those of AGNs.
Many properties of IRS 7B, e.g. the large equivalent width of the fluorescent
X-ray iron line and the spectral energy distribution,
suggest it to be a Class 0 source or an object between Class 0 and Class I
phase (\cite{hamaguchi2005}; \cite{nutter2005}; \cite{groppi2007}). 

Given the spectral indices and other properties of IRS 7B, the continuum 
radiation is probably coming from the surrounding gas which is shock-ionized 
by the high velocity jet. The radio lobes associated with the thermal  
radio jet possible have a gyrosynchrotron emission component. This could be 
understood in terms of DSA of electrons (see Sect. 2.1). 

\subsubsection{Brown 5 - a brown dwarf ?}

This is the Radio Source 5 detected by Brown (1987) 
using the VLA at 6 cm. It has been suggested that the infrared counterpart of 
Brown 5 could be a brown dwarf if it lies in the R CrA cloud 
\textbf{(\cite{wilking1997}; \cite{feigelson1998}; see also
\cite{forbrich2006})}.
The infrared source has been detected in the $H$ and $K$ bands only,
with $K=16.4$, and $H-K=1.45$. The value of $H-K$ suggests a large
interstellar or circumstellar reddening ($(H-K)_0 \sim 0.5$ for an M9V
dwarf, \cite{wilking1999}). Therefore the object is likely to lie in
the cloud or behind it. It should be noted that brown dwarfs are seldom
detected in radio (see, e.g., \cite{gudel2002}; \cite{osten2006}).
During flares the flux densities of nearby ($d \sim 10$ pc)
brown dwarfs can increase to a few millijanskys, but the average value is
an order of magnitude lower
(radio luminosity $L_{\nu, \, 5 \, {\rm GHz}} < 10^{15}$
ergs s$^{-1}$ Hz$^{-1}$). Assuming that the object lies at the distance
of the CrA cloud its (nearly constant) radio luminosity is
$L_{\nu, \, 5 \, {\rm GHz}} \sim  3 \cdot 10^{16}$ ergs s$^{-1}$ Hz$^{-1}$.
This high luminosity would be characteristic of a magnetically active star
rather than a brown dwarf (\cite{gudel2002}). On the other hand, the fact
that Brown 5 has not been detected in X-rays (\cite{forbrich2006};
\cite{forbrich2007}) makes coronal emission from a late-type star a less
likely alternative.

The fluxes observed here are similar at all wavelengths to those 
observed by Suters et al. (1996), Feigelson et al. (1998) 
and Forbrich et al. (2006). 
On one day Suters et al. detected an unusually high 3 cm flux, 
and later on, Forbrich et al. (2006) found that the source is variable 
on timescales of days to months. 
 
The radio spectral indices, $\alpha_{3 \, \rm cm}^{6 \, \rm cm}=-0.34\pm0.12$  
and $\alpha_{6 \, \rm cm}^{20 \, \rm cm}=-0.30\pm0.01$, 
indicate optically thin gyrosynchrotron emission, 
and so magnetic field activity in this object. 
The former spectral index is similar to that found by Suters et al.  
on Jul 12, 1992 ($-0.28\pm0.30$), but our 6/20 cm index is very different from 
their values ($\alpha_{6 \, \rm cm}^{20 \, \rm cm} \sim0.1$).
Spectral index derived from 2000 May 31 / Jun 9 data, 
$\alpha_{3 \, \rm cm}^{6 \, \rm cm}=0.76\pm0.29$, is probably not meaningful 
since the source is variable with a scale of days. 

According to Suters et al. (1996) the high obscuration and rapid variations 
make it more likely that Brown 5 is a PMS star rather than an extragalactic
source. 
However, extragalactic nature cannot be totally ruled out as interstellar 
scintillation can cause so called IntraDay Variability 
(IDV, see \cite{gabanyi2007}). 

\subsubsection{IRS 5} 

The Class I protostar IRS 5 is known to be a highly variable 
radio source. 
Our ATCA observations yield mainly similar flux densities  
to those observed by Brown (1987), Suters et al. (1996), Feigelson et al. 
(1998) and Forbrich et al. (2006; 2007). At 20 cm we found a flux of 
0.40 mJy, over 2 times lower than the lowest value found by Suters et al. 
(1996). Suters et al. detected a flare of 6.6 mJy at 6 cm in 1992 and  
concluded that IRS 5 experienced an outburts in mass loss.  
The radio spectral indices of IRS 5 found in the 
present study are $\alpha_{3 \, \rm cm}^{6 \, \rm cm}=-0.12\pm0.07$ and 
$0.26\pm0.16$ and $\alpha_{6 \, \rm cm}^{20 \, \rm cm}=0.54\pm0.08$. 
We note that the latter two values can be deceptive since 
the source is highly variable. Suters et al. found spectral indices  
$\alpha_{3 \, \rm cm}^{6 \, \rm cm} \sim0.1$ and $\sim0.9$, and  
$\alpha_{6 \, \rm cm}^{20 \, \rm cm}$ varied from $\sim -0.01$ to $-0.1$,  
all very different from those derived here. On the other hand, the indices 
of Suters et al. have rather large errors. 

IRS 5 has been detected in X-rays (Koyama et al. 1996; 
\cite{neuhauser1997}), and recently Forbrich et al. (2006; 2007) 
found it to be the most variable source in X-rays in the R CrA region 
(see also \cite{forbrich2007b}).   
Feigelson et al. (1998) detected circularly polarised non-thermal emission 
from IRS 5 at 3.5 cm, which is also the first such detection for a YSO. 
This was later confirmed by Forbrich et al. (2006),  
but interestingly, there was also occasion when IRS 5 did not show circular 
polarisation in their observations. In the present study, the Stokes-$V$ 
emission of IRS 5 was detected only at 3 cm on May 31, 2000, with the flux 
density of $0.23\pm0.04$ mJy, the polarisation fraction, $V/I$, being as 
large as about 33 \%. The \textbf{$3\sigma$} upper limits of $V$ emission at
3 and 6 cm observed in 1998 Jan 9/10, at 3 cm in 2000 May 28,
and at 6 and 20 cm in 2000 Jun 9/10 are about 0.07, 0.20, 0.18, 0.06,
and 0.04 mJy, respectively. 
The corresponding upper limits of $V/I$ are about 9 and 19 \% at
3 and 6 cm in 1998 Jan 9/10, respectively, and about 10 \% during other
measurements in 2000.

The slightly negative spectral index between 3 and 6 cm,
and circular polarisation suggest that the radio emission has
a substantial non-thermal gyrosynchrotron component. This conforms with the
fact that IRS 5 does not appear to be powering an outflow
(\cite{feigelson1998}; see also \cite{wang2004}).
Detectable non-thermal emission implies that the immediate surroundings of the
star are relatively free of optically thick ionized gas (see Sect. 2.2).
The classification of IRS 5 is complicated by the fact that it is a binary
with a separation of $\sim0\farcs6$ or 102 AU at 170 pc 
(\textbf{\cite{chen1993}}; \cite{nisini2005}). The binary is also
marginally resolved in X-rays (\cite{forbrich2007b}).
This together with the rapid variability make the determination of the spectral
indices uncertain. Nevertheless, the radio properties of this source suggest
that at least one of the components has advanced beyond the Class I stage.

\subsubsection{Other detections} 

The T Tauri star \textbf{IRS 6} was detected only at 6 cm in the 1998 
visibility data with the flux density $0.46 \pm 0.04$ mJy. Thus, the spectral
index cannot be determined. The $3\sigma$ upper limits of
the 3 cm flux density are 0.02 and 0.03 mJy (on Jan 9/10, 1998,
and May 28/31, 2000, respectively), and the $3\sigma$ upper limits of
the 6 and 20 cm flux densities are 0.13 and 0.10 mJy (on Jun 9/10, 2000).
There are no reported radio emission from this source before 
Forbrich et al. (2006). Forbrich et al. (2006; 2007) also found very 
low fluxes of $\sim0.1$ mJy at 3.6 cm. 
 
\textbf{IRS 1} is another Class I object in the \textit{Coronet} cluster 
(\cite{wilking1986}; \cite{wilking1992}). 
It is also known as HH 100-IR, and is likely a driving source of the 
HH flow consisting of HH 96, 97, 100, and 98.  
The flux densities of IRS 1 in the present work are very similar to those 
observed by Brown (1987), Suters et al. (1996), Feigelson et al. (1998) and 
Forbrich et al. (2006). At 20 cm IRS 1 remained below the $3\sigma$ level 
of 0.36 mJy. 
The spectral index $\alpha_{3 \, \rm cm}^{6 \, \rm cm}=1.96\pm0.32$  
is consistent with the value $\alpha_{3 \, \rm cm}^{6 \, \rm cm}=1.67\pm0.72$
derived by Suters et al. (1996). This high spectral index suggests optically 
thick free-free emission, which can attributed to shocked material. 
IRS 1 has been detected as a hard X-ray source (Koyama et al. 1996; 
\cite{neuhauser1997}). 
Even though IRS 1 does not show significant radio variability, it shows 
clear signs of X-ray variability (\cite{suters1996}; \cite{forbrich2006}). 
According to Nisini et al. (2005) a very high fraction of the bolometric 
luminosity of IRS 1 is due to accretion. In view of these results it seems 
possible that the cm emission originates in an accretion
shock in the circumstellar disk. 

The Herbig Ae star \textbf{R CrA} has remained undetected in several surveys, 
indicating long-term variability (see \cite{forbrich2006} and references
therein). 
Our 3 cm flux densities are somewhat higher than those observed by  
Forbrich et al. (2006; 2007). The 6 cm flux of R CrA was $0.29\pm0.03$ mJy.  
The source was not detected at 20 cm above the $3\sigma$ level of 
0.36 mJy.

Koyama et al. (1996) detected hard X-ray emission close to the position of 
R CrA and Forbrich et al. (2006) clearly detected X-ray emission from R CrA 
in all of their \textit{Chandra} and XMM-\textit{Newton} datasets. 
As Forbrich et al. (2006) pointed out, the X-ray emission from this star does 
not fit into our understanding of stellar X-ray emission mechanisms 
(magnetically driven corona or shocks caused by strong stellar wind; 
see also \cite{forbrich2007b}). 
 
Given the substantial error in the spectral index, 
$\alpha_{3 \, \rm cm}^{6 \, \rm cm}=0.41\pm0.33$, it is not 
possible to make firm conclusions about the emission mechanism for this 
continuum radiation. However, since R CrA should not drive very strong 
stellar wind, 
the upper limit of the spectral index ($\sim0.7$) is 
excluded. On the other hand, since R CrA is likely to be a driving source 
of collimated outflow (see, e.g., \cite{suters1996}), radio emission is 
probably produced by a collimated, ionized flow. 
 
\subsubsection{Non-detections} 
 
\textbf{IRS 2}: The Class I protostar IRS 2 is one of the strongest 
X-ray sources among the objects studied by Forbrich et al. (2006). 
The non-detection is probably due to primary beam attenuation far from 
the observation's phase centre. 
 
\textbf{IRS 9}: The Class I protostar IRS 9 remains undetected at radio 
also in previous radio observations (see \cite{forbrich2006}). 
 
\textbf{T CrA}: The Herbig Ae star, T CrA, of spectral F0e is one of 
the most massive stars in the R CrA core.
The non-detection is consistent with the earlier studies of Brown (1987)
and Suters et al. (1996).

\section{Summary and conclusions} 

We have conducted a survey of radio continuum emission at 3, 6 and 20 
cm in the R CrA star-forming region using the ATCA.  Altogether eight 
pointlike sources were detected. Seven of these can be assigned to 
YSOs, having counterparts in the X-rays, infrared or submm. We 
studied their spectral indices, variability, and polarisation to gain 
information on possible emission mechanisms. In this manner we 
examined the classifications of the YSOs in the light of the 
evolutionary scenario presented by Gibb et al. (1999), which relates 
radio properties to protostellar stages. 

Probably the youngest protostars in our sample are Brown 9 and IRS 7B.
Brown 9 is a Class 0 candidate and IRS 7B has been suggested to represent
a transition stage between Class 0 and Class I (\cite{groppi2007}).
The spectral indices of these two sources are
consistent with free-free emission from a collimated jet.
IRS 7B shows signs of extended lobes at 6 
and 20 cm. The lobes may have a gyrosynchrotron component which possibly 
can be understood in terms of Fermi acceleration of electrons in shocks 
associated with high velocity jets from the protostar. 

The three Class I protostars detected in this survey, IRS 7A, IRS 5 
and IRS 1 have very different characteristics. The spectra of IRS 1 
and IRS 7A are consistent with free-free emission, with the 
distinction that emission from IRS 1 is optically thick, possibly 
originating in shocks, whereas emission from IRS 7A is optically thin, 
suggesting a collimated jet. Radio emission from the wide binary star 
IRS 5 has a substantial non-thermal component indicated by a negative 
spectral index, rapid variability, and strong Stokes-$V$ emission on one 
of days of observation. These characteristics are signs of magnetic 
activity close to the stellar photosphere. This suggests that at least 
one of the components of IRS 5 is free of absorbing ionized gas, and 
may have already reached an advanced stage. 

Two infared-bright Class I candidates in the region, 
IRS 2 and IRS 9, remained undetected in radio.
In accordance with previous results, R CrA (HAe) and
the Class II star IRS 6 (CTTS) show weak radio emission.
The so far unclassified bright radio source Brown 5 has a mildly negative 
spectral index, characteristic of optically thin gyrosynchrotron 
radiation. This source is likely to be an obscured Class III star 
or an extragalactic source.

The present observations are in rough agreement with the scheme of 
Gibb et al. (1999) where the dominant emission mechanism changes with 
the age of an YSO. It was found, however, that Class I stars form a 
very heterogeneous group as to their radio properties. This can be 
perhaps understood in terms of a rapid decline in the jet activity, 
which also changes the efficiency of free-free absorption. 
The results from IRS 7B suggest that energetic jets from
very young protostars can give rise to non-thermal emission via
shock acceleration.

The experiences from the R CrA stellar cluster have shown that 
observations in a wide range of wavelengths, in fact from X-rays to 
radio, are required to understand the nature of the youngest 
protostars (\cite{forbrich2007}; \cite{groppi2007}). At the same it 
is evident that in the radio regime, higher sensitivity and angular 
resolution than available now are needed to detect a significant 
fraction of YSOs in a dense cluster like R CrA, and to study the 
physical processes which determine their evolution. 

\begin{acknowledgements}
   
We thank the ATCA staff for their help during 
the observations, and the referee for very helpful comments 
and suggestions. The Helsinki group acknowledges support from the 
Academy of Finland through grants 1117206 and 1210518. 

\end{acknowledgements}

\end{document}